\documentclass[a4paper,11pt]{article}

\newcommand{\sect}[1]{\setcounter{equation}{0}\section{#1}}

\usepackage{amssymb,latexsym}

\begin{document}
\topmargin -10mm
\oddsidemargin 0mm

\renewcommand{\thefootnote}{\fnsymbol{footnote}}
\newcommand{\nn}{\nonumber\\}
\begin{titlepage}
\begin{flushright}
OU-HET 550 \\
hep-th/0601044
\end{flushright}

\vspace{5mm}
\begin{center}
{\Large \bf  Holography and D3-branes in Melvin Universes}
\vspace{12mm}

{\large
Rong-Gen Cai~\footnote{Email address: cairg@itp.ac.cn }}\\
\vspace{8mm} { \em Institute of Theoretical Physics, Chinese
Academy of Sciences,\\
 P.O. Box 2735, Beijing 100080, China}
\vspace{10mm}

{\large Nobuyoshi Ohta~\footnote{Email address:
ohta@phys.sci.osaka-u.ac.jp.  Address after 31 March 2006:
Department of Physics, Kinki University,
Higashi-Osaka, Osaka 577-8502,Japan}}\\
\vspace{8mm} { \em Department of Physics, Osaka University,
Toyonaka, Osaka 560-0043, Japan}

\end{center}
\vspace{10mm} \centerline{{\bf{Abstract}}}
 \vspace{5mm}
 Recently,in a paper hep-th/0511197, it was found that non-commutative super
Yang-Mills (NCSYM) theory with space-dependent noncommutativity
can be formulated as a decoupling limit of open strings ending on
D3-branes wrapping a Melvin universe supported by a flux of the
NSNS B-field. Under S-duality, we show that this theory turns into
a noncommutative open string (NCOS) theory with space-dependent
space-time noncommutativity and effective space-dependent string
scale. It is an NCOS theory with both space-dependent space-space
and space-time noncommutativities under more general
$SL(2,\mathbb{Z})$ transformation. These space-dependent
noncommutative theories (NCSYM and NCOS) have completely the same
thermodynamics as that of ordinary super YM theory, NCSYM and NCOS
theories with constant noncommutativity in the dual supergravity
description. Starting from black D3-brane solution in the Melvin
universe and making a Lorentz boost along one of spatial
directions on the worldvolume of D3-branes, we show that the
decoupled theory is a light-like NCSYM theory with space-dependent
noncommutativity in a static frame or in an infinite-momentum
frame depending on whether there is a gravitational pp-wave on the
worldvolume of the D3-branes.

\end{titlepage}

\newpage
\renewcommand{\thefootnote}{\arabic{footnote}}
\setcounter{footnote}{0} \setcounter{page}{2}

%%=====================section 1====================
\sect{Introduction}

Over the past decade one of the most important progresses
in string/M theories is that some super Yang-Mills (SYM)
theories can be naturally realized as a decoupling limit of open
strings ending on D-branes and they have dual supergravity
descriptions~\cite{Mald1,Itzh,Gubs,Witten1}. As a concrete
example, ${\cal N}=4$ SYM theory in four dimensions
is the decoupling limit of open string theory on D3-branes, and
its dual supergravity geometry is $AdS_5 \times S^5$, which is
near horizon geometry of D3-brane solution in Type IIB
supergravity. Furthermore, thermodynamics of black D3-branes in the
decoupling limit can be identified with that of the ${\cal N}=4$
SYM theory~\cite{Gubs2,Witten2}.

The ${\cal N}=4$ SYM theory can be deformed in various manners.
One of them is the one by adding a dimension 6
operator~\cite{operator6}, its bosonic part is
\begin{equation}
\label{1eq1}
{\cal O}_{\mu\nu}=\frac{1}{2g^2_{\rm YM}}{\rm Tr}\left(
F_{\mu\sigma}F^{\sigma\rho}F_{\rho\nu}-F_{\mu\nu}F_{\rho\sigma}F^{\rho\sigma}
+2 F_{\mu\rho}\Sigma_{i=1}^6\partial_{\nu}\phi^i\partial^{\rho}\phi^i
-\frac{1}{2}F_{\mu\nu}\Sigma_{i=1}^6\partial_{\rho}\phi^i
\partial^{\rho}\phi^i \right ),
\end{equation}
where $g_{\rm YM}$ is the SYM coupling constant, $F_{\mu\nu}$ is
the $U(N)$ field strength and $\phi^i$ ($i=1,2,\cdots,6$) are the
adjoint scalars. This deformed theory can be extended to a
complete theory, noncommutative SYM theory. From the point of view
of open string theory on the D3-branes, the deformed theory
corresponds to the decoupling limit of open string theory on the
D3-branes with nonzero NSNS B-field. Indeed, four-dimensional
NCSYM theory with a constant space-space noncommutativity
parameter can be realized as a decoupling limit of open strings
ending on D3-branes with a constant space-like NSNS B-field (or
constant magnetic field) on the worldvolume~\cite{NCYM}. Its dual
supergravity  description is the near horizon geometry of (D3, D1)
bound states~\cite{GNCYM}. Under S-duality, the bound state
(D3, D1) turns out to be (D3, F1) bound state. It was found that
under the S-duality, the four-dimensional NCSYM theory with a
constant space-space noncommutativity changes to noncommutative
open string (NCOS) theory with a constant time-space
noncommutativity parameter, rather than an NCSYM theory with a
time-space noncommutativity~\cite{NCOS}. This NCOS theory is a
decoupling limit of open strings ending on D3-branes with a
constant time-like NSNS B-field (or constant electric field)  on
the worldvolume. Note that there is an $SL(2,\mathbb{Z})$ symmetry in type
IIB string theory. One can have more general NCOS theory with
constant time-space and space-space noncommutativity parameters,
and its dual supergravity description is the geometry of the bound
state (D3, (D1, F1)) solutions~\cite{DNCOS}.

While NCSYM theories with space-space noncommutativity is a
well-defined theory, an NCSYM theory with a time-space
noncommutativity will violate the unitarity~\cite{unitary}.
However, it is interesting to note that an NCSYM theory with
light-like noncommutativity is well-defined and obeys the
unitarity~\cite{unitarity}. Supergravity descriptions dual to
light-like NCSYM theories are investigated in~\cite{Alih,CO1}.
Note that the background, on which NCSYM and NCOS theories reside,
is the flat Minkowski spacetime, and the noncommutativity
parameters are constant. One can consider D-branes in
time-dependent spacetime, and obtain NCSYM theories in
time-dependent background~\cite{Hasi1,Alih2}. In this case, after
an S-duality, the resulting theory is an NCOS theory also in a
time-dependent background but now with  time-dependent time-space
noncommutativity parameter and  time-dependent string
scale~\cite{CLO}.

More recently, considering D3-branes wrapping a Melvin universe
supported by a flux of the NSNS B-field, Hashimoto and
Thomas~\cite{HT} have shown that the resulting theory on the
worldvolume in a decoupling limit is a four-dimensional NCSYM
theory with a space-dependent noncommutativity parameter and the
supersymmetry is completely broken. Along the same line, some
generalizations to other dimensions and in M theory have been
investigated in \cite{ASY}. These works widen the family of
nonlocal field theories and provide more examples to realize the
gravity/gauge field correspondence.

In this note we will further study D3-branes in Melvin universes,
and generalize the works \cite{HT,ASY} in some aspects. In the
next section we will first obtain black D3-brane solutions in
Melvin universes and study its thermodynamics, and find that in
the dual supergravity description, the thermodynamics of NCSYM
with space-dependent noncommutativity is completely identical to
that of NCSYM theory with constant noncommutativity and of
ordinary SYM theory.  In Sec.~3 we study the S-duality of the
NCSYM theory and find that the resulting theory is an NCOS theory
with space-dependent time-space noncommutativity  and
space-dependent string scale. More general $SL(2,\mathbb{Z})$
transformation gives an NCOS theory with both space-dependent
time-space and space-space noncommutativities.  In Sec.~4 we make
a Lorentz boost along one of worldvolume directions of D3-branes
in Melvin universe, and obtain a light-like NCSYM theory with a
space-dependent light-like noncommutativity parameter. We end this
paper with our conclusion in Sec.~5.

\sect{NCSYM and Black D3-branes in Melvin Universe}

To construct the black D3-brane solution in the Melvin universe,
we follow the steps given by Hashimoto and Thomas~\cite{HT}: (1)
starting from ordinary black D3-brane solution in an
asymptotically flat spacetime, and making a T-duality along one of
spatial directions, say $z$, on the worldvolume of D3-branes (The
other two spatial coordinates denoted by $x$ and $y$, or $r$ and
$\varphi$), one obtains the black D2-brane solution with a smeared
transverse coordinate $z$; (2) Making a twist, $d\varphi \to
d\varphi +\eta dz$ (where $\eta$ is a constant), and T-dualizing
along the coordinate $z$, we get the black D3-brane solution in a
Melvin universe~\footnote{Here $F_5=*F_5$ should be understood.}
\begin{eqnarray}
\label{2eq1} && ds^2= H^{-1/2}[-fdt^2 +dr^2 +h(r^2 d\varphi^2
+dz^2)]+H^{1/2}(f^{-1}d\rho^2 +\rho^2d\Omega_5^2), \nn
 &&
e^{2\phi}= g^2_s h, \ \ \ 2\pi \alpha'B = \frac{r^2
\eta}{H+r^2\eta^2}d\varphi \wedge dz, \nn
 &&
A_2=g_s^{-1} H^{-1} \eta \ r \coth \alpha \ dt\wedge dr,
 \nn
 && A_4=g_s^{-1}H^{-1} r h\coth \alpha \ dt \wedge dr
\wedge d\varphi \wedge dz,
\end{eqnarray}
where $g_s$ is the closed string coupling constant at infinity,
\begin{equation}
\label{2eq5}
f= 1-\frac{\rho_0^4}{\rho^4}, \ \ \  H= 1+\frac{\rho_0^4
\sinh^2\alpha}{\rho^4},\ \ \ h=\frac{H}{H+\eta^2r^2},
\end{equation}
$\rho_0$ is the Schwarzschild  parameter and
$\alpha$ is the D3-brane charge parameter. When $\rho_0 \to 0$,
$\alpha \to \infty$, but keeping
$\rho_0^4 \sinh\alpha \cosh\alpha =\mbox{const.} = 4 \pi g_s N\alpha'^2$ with
$N$ being the number of D3-branes, we obtain D3-brane solution
in a Melvin universe.

The black D3-brane solution (\ref{2eq1}) has a horizon at $\rho=\rho_0$. The
associated Hawking temperature and entropy are
\begin{equation}
\label{2eq6}
T= \frac{1}{\pi \rho_0 \cosh\alpha}, \ \ \  S= \frac{\pi^3 V_3}{4
G_{10}} \rho_0^5\cosh\alpha,
\end{equation}
respectively, where $G_{10}= 2^3\pi^6 g_s^2 \alpha'^4 $ is the
Newtonian constant in ten dimensions and $V_3$ is the spatial
volume of D3-brane worldvolume.  The solution (\ref{2eq1}) is not
asymptotically flat. The usual approach does not work for getting
the mass of the solution. Taking the case with $\rho_0=0$ as a
reference background, we obtain the mass of the solution by using
the background subtraction approach
\begin{equation}
\label{2eq7}
 M=\frac{5\pi^2\rho_0^4 V_3}{16
 G_{10}}\left(1+\frac{4}{5}\sinh^2\alpha \right).
\end{equation}
We note that these thermodynamic quantities are completely the same as
those of black D3-brane solutions with/without NSNS
B-field~\cite{GNCYM}.

To discuss the low-energy decoupling limit of open string theory
in the background of the D3-branes in the Melvin universe
(\ref{2eq1}), a good starting point is still the Seiberg-Witten
relation connecting the open string and closed string moduli~\cite{SW}
\begin{eqnarray}
\label{2eq8}
&& G_{ij}=g_{ij} -(2\pi \alpha')^2(Bg^{-1}B)_{ij},
  \nonumber \\
&& \Theta^{ij}=2\pi \alpha' \left(\frac{1}{g+2\pi
\alpha'B}\right)^{ij}_{A}, \nonumber \\
&& G^{ij}=\left(\frac{1}{g+2\pi \alpha'B}\right)^{ij}_{S}, \nonumber \\
&& G_s=g_s\left(\frac{\det G_{ij}}{\det(g_{ij}+2\pi
\alpha' B_{ij})}\right)^{1/2},
\end{eqnarray}
where $(~)_A$ and $(~)_S$ stand for the antisymmetric and
symmetric parts, respectively.  The open string moduli occur in
the disk correlators on the open string worksheet boundaries
\begin{equation}
< X^i(\tau)X^j(0)>
=-\alpha' G^{ij} \ln (\tau)^2
+\frac{i}{2}\Theta^{ij}\epsilon(\tau),
\end{equation}
where $\epsilon (\tau)$ is a function that is $1$ or $-1$ for
positive or negative $\tau$.
For the black D3-brane solution (\ref{2eq1}) %in the Mevlin universe,
in the limit of the Melvin universe, that is, $H=f=1$,
we can examine the open string moduli by using Seiberg-Witten
relation (\ref{2eq8}). We find that the open string metric
$G^{ij}$ is given by
\begin{equation}
\label{2eq10}
G_{ij}dx^i dx^j =-dt^2 + dr^2 +r^2 d\varphi^2 +dz^2,
\end{equation}
the nonvanishing noncommutativity parameter is
\begin{equation}
\label{2eq11}
\Theta^{\varphi z}= -2\pi \alpha' \eta,
\end{equation}
and the open string coupling constant $G_s=g_s$. In the decoupling
limit
\begin{equation}
\label{2eq12}
\alpha' \to 0: g_s =\mbox{const.}, \ \  \eta = \triangle /\alpha',
\end{equation}
where $\triangle $ is constant, one can see from (\ref{2eq10}) and
(\ref{2eq11}) that open string
moduli are well-defined~\cite{HT2,HT}. The decoupling limit of the
open string theory is a four-dimensional NCSYM theory residing on
the spacetime (\ref{2eq10}) with YM coupling $g_{\rm YM}^2= 2\pi
g_s$ and noncommutativity parameter
\begin{equation}
\label{2eq13} \Theta^{\varphi z}= -2\pi \triangle.
\end{equation}
Though it appears that the noncommutativity parameter is constant in the
coordinates (\ref{2eq10}), it is space-dependent in the coordinates
$ds^2= -dt^2+dx^2 +dy^2 +dz^2$:
\begin{equation}
\label{2eq14}
\Theta^{xz}=-2\pi y \triangle , \ \ \
\Theta^{yz}=2\pi x \triangle.
\end{equation}
This noncommutativity parameter is divergence free and obeys the
Jacobi identity~\cite{HT}. Since the noncommutativity parameter is
space-dependent, to construct a gauge invariant action in such a
background, the product between operators has to use the so-called
Kontsevich product~\cite{Kont}
\begin{equation}
f*g =fg +i\theta^{ij} \partial_i f \partial_j g -\frac{1}{8}\theta
^{ij}\theta^{kl}\partial_{ik}f\partial_j\partial_l g
-\frac{1}{12}\theta^{ij}\partial_j \theta^{lm}
(\partial_{i}\partial_l f\partial_m g -\partial_l f
\partial_i\partial_m g) +{\cal O}(\theta^3).
\end{equation}
When $\theta$ are constants, the Kontsevich product reduces to the
Moyal product.

The dual supergravity  description of the NCSYM theory with
space-dependent noncommutativity parameter can be obtained from
the black D3-brane solution (\ref{2eq1}) by taking the decoupling
limit (\ref{2eq12}) combined with
\begin{equation}
\label{2eq16}
\rho= \alpha' u,  \ \ \ \rho_0 =\alpha' u_0.
\end{equation}
The supergravity solution becomes
\begin{eqnarray}
\label{2eq17}
&& ds^2 = \alpha'\left \{\frac{u^2}{R^2}(-\tilde f dt^2 +dr^2
+\tilde h (r^2 d\varphi^2 +dz^2)) + \frac{R^2}{u^2} (\tilde
f^{-1}du^2 +u^2 d\Omega_5^2)\right \},
\nonumber \\
&& e^{2\phi}= g_s^2 \tilde h, \ \ \ 2\pi \alpha' B = \alpha' r^2
\triangle u^4 R^4 \tilde h d\varphi \wedge dz, \nonumber \\
&& A_2= \alpha' g_s^{-1} \triangle \, r \frac{u^4}{R^4} dt\wedge
dr, \nonumber \\
&& A_4 = \alpha'^2 g^{-1}_s r \tilde h \frac{u^4}{R^4}dt\wedge
dr\wedge d\varphi \wedge dz,
\end{eqnarray}
where
\begin{equation}
\label{2eq18}
\tilde f = 1-\frac{u_0^4}{u^4}, \ \ \ \tilde h= \frac{R^4}{R^4 +
r^2\triangle ^2 u^4},
\end{equation}
and $R^4= 4\pi g_s N = 2 g_{\rm YM}^2 N$. When $u_0=0$, namely,
$\tilde f=1$, our solution (\ref{2eq17}) reduces to the one
given in \cite{HT2,HT}. In other words, our solution
(\ref{2eq17}) describes the gravity dual of the NCSYM theory at
finite temperature
\begin{equation}
\label{2eq19}
T= \frac{u_0}{\pi R^2}.
\end{equation}
{}From the solution (\ref{2eq17}) we can see that the solution
deviates from the supergravity dual $AdS_5\times S^5$ of the
ordinary SYM theory by a factor $\tilde h$. When $u^4 \ll
R^4/r^2\triangle^2$, we see from (\ref{2eq18}) that the
deviation is small. Since the coordinate $u$ stands for the
energy scale of dual YM theory in the AdS/CFT correspondence,
the NCSYM theory with space-dependent noncommutativity has
only small deviation from the ordinary SYM theory in the low-energy
limit as the case of NCSYM with constant noncommutativity
parameter~\cite{GNCYM}. The NCSYM theory at the temperature
(\ref{2eq19}) has entropy
\begin{equation}
\label{2eq20} S =\frac{R^2 u_0^3 V_3}{2^5 \pi^3
g_s^2}=\frac{1}{2}\pi^2 N^2 T^3V_3.
\end{equation}
Once again, the entropy has completely the same expression as the
one for the ordinary SYM theory and NCSYM with a constant
noncommutativity parameter in the supergravity duals~\cite{GNCYM}.
This indicates that in the large $N$ limit, even for the
space-dependent noncommutativity case, the total number of degrees
of freedom of NCSYM theories coincides with the one for ordinary
SYM theory for any energy scale~\cite{NCYM}.

\sect{S-duality and NCOS with space-dependent Noncommutativity}

It is shown that the S-duality of NCSYM theory with constant
space-space noncommutativity results in an NCOS theory with
constant time-space noncommutativity~\cite{NCOS}. This also holds
for time-dependent background~\cite{CLO}. Therefore one may expect
that S-duality for the NCSYM theory with space-dependent
noncommutativity will give an NCOS theory with space-dependent
time-space noncommutativity. In this section we will show that it
is indeed the case. Furthermore more general $SL(2,\mathbb{Z})$
transformation will result in an NCOS theory with both
space-dependent time-space and space-space noncommutativities.

Making an S-duality along the coordinate $z$ for the black
D3-brane solution (\ref{2eq1}), we obtain~\footnote{Here we have
rescaled the closed string coupling constant as $1/g_s \to g_s$.}
\begin{eqnarray}
\label{3eq1} && ds^2= H^{-1/2}h^{-1/2}(-fdt^2 +dr^2 +h(r^2
d\varphi^2 +dz^2))+H^{1/2}h^{-1/2}(f^{-1}d\rho^2
+\rho^2d\Omega_5^2), \nn \label{3eq2} && e^{2\phi}= g^2_s h^{-1},
\ \ \ 2\pi \alpha'B=H^{-1} \eta \ r \coth \alpha \ dt\wedge dr,
\nn \label{3eq3} && A_2=g_s^{-1} \frac{r^2
\eta}{H+r^2\eta^2}d\varphi \wedge dz, ~~~ \label{3eq4}
A_4=g_s^{-1}H^{-1} r h \coth \alpha \ dt \wedge dr \wedge d\varphi
\wedge dz.
\end{eqnarray}
To see what a well-defined decoupled theory can be obtained for
open string theory in the background (\ref{3eq1}), we use again
the Seiberg-Witten relation (\ref{2eq8}) for the solution
(\ref{3eq1}) in the ``flat" limit, namely, $f=H=1$. We find that
the open string metric is
\begin{equation}
\label{3eq5} G_{ij} dx^i dx^j = h^{1/2}(-dt^2 +dr^2 +r^2d\varphi^2 +dz^2),
\end{equation}
the nonvanishing noncommutativity parameter
\begin{equation}
\label{3eq6} \Theta^{tr}=2\pi \alpha' \eta r,
\end{equation}
and the open string coupling constant is still $G_s=g_s$. In the
decoupling limit (\ref{2eq12}),  a set of well-defined open string
moduli can be obtained as follows:
\begin{eqnarray}
\label{3eq7}
&& \alpha'G^{ij} \partial_i \partial_j = \triangle r (d^2/d t^2 +d^2/d r^2
 + d^2/r^2 d\varphi^2 +d^2/d z^2),
\\
\label{3eq8}
&& \Theta^{tr}= 2\pi \triangle r.
\end{eqnarray}
Indeed we see that the resulting theory is an open string theory
with coupling constant $G_s=g_s$ and space-dependent time-space
noncommutativity parameter $\Theta^{tr}$. The effective open string scale
is $\alpha'_{\rm eff}= r\triangle$, having a space dependence.
As usual, the supergravity dual
of the NCOS theory can be found from the solution (\ref{3eq1}) in
the decoupling limit (\ref{2eq12}) and (\ref{2eq16})
\begin{eqnarray}
\label{3eq9}
&& ds^2 =\alpha'\tilde h^{-1/2}\left \{\frac{u^2}{R^2}(-\tilde f dt^2 +dr^2
+\tilde h (r^2 d\varphi^2 +dz^2)) + \frac{R^2}{u^2} (\tilde
f^{-1}du^2 +u^2 d\Omega_5^2)\right \}, \nonumber \\
&& e^{2\phi}= g_s^2 \tilde h^{-1}, \ \ \ A_2 = \alpha'g_s^{-1} r^2
\triangle u^4 R^4 \tilde h d\varphi \wedge dz, \nonumber \\
&& 2\pi \alpha' B = \alpha'  \triangle \, r \frac{u^4}{R^4} dt\wedge
dr, \nonumber \\
&& A_4 = \alpha'^2 g^{-1}_s r \tilde h \frac{u^4}{R^4}dt\wedge
dr\wedge d\varphi \wedge dz.
\end{eqnarray}
It is easy to show that the temperature and entropy of the
supergravity dual (\ref{3eq9}) are still given by (\ref{2eq19})
and (\ref{2eq20}), respectively. This implies that the NCOS theory
has the same thermodynamics and degrees of freedom as those of the
NCSYM theory with space-dependent noncommutativity described by
(\ref{2eq17}).

It is well known that type IIB string theory has an $SL(2,\mathbb{Z})$
invariance. Considering the $SL(2,\mathbb{Z})$ transformation to the
solution (\ref{2eq1}), we can get more general black D3-brane
solution in Melvin universe.  Since the axion field vanishes for
the solution (\ref{2eq1}), we have the transformed solution
\begin{equation}
\label{3eq10} \tilde \tau = \frac{a \tau +b}{c\tau +d}, \ \ \
\tau =i e^{-\phi}, \ \ \ ad-bc =1, \ \ \ a,b,c,d \in \mathbb{Z},
\end{equation}
whose imaginary part gives the transformed dilaton field
\begin{equation}
\label{3eq11}
e^{-\tilde \phi}= \frac{e^{-\phi}}{|c\tau +d|^2},
\end{equation}
and real part gives a nonvanishing axion field $\tilde \chi$.
The new configuration has the form~\footnote{Here we have made
rescaling of coordinates and $\eta$ as follows: $(t,r,z,
\rho, 1/\eta) \to g_s^{1/2}/(c^2+d^2g_s^2)^{1/4} \times (t,r,z,
\rho, 1/\eta) $.}
\begin{eqnarray}
\label{3eq12}
&& ds^2= H^{-1/2}h^{-1/2}\sqrt{\frac{c^2+d^2g_s^2 h}{c^2+d^2
g_s^2}}\left [-fdt^2 +dr^2 +h(r^2d\varphi^2+dz^2)+H(f^{-1}d\rho^2
+\rho^2 d\Omega_5^2)\right], \nonumber
\\
&& e^{-2\tilde \phi}= g_s^{2} h (c^2+d^2g_s^2 h)^{-2}, \ \ \
\tilde \chi= \frac{ac+db g_s^2h}{c^2+d^2g_s^2h}, \nonumber \\
&& 2\pi \alpha' \tilde B=
\frac{r^2\eta}{H+r^2\eta^2}\frac{dg_s}{\sqrt{c^2+d^2g_s^2}}d\varphi\wedge
dz -\frac{\eta r \coth \alpha}{H}\frac{c}{\sqrt{c^2+d^2g_s^2}}dt\wedge
dr, \nonumber \\
&& \tilde A_2= H^{-1} \eta r \coth\alpha
\frac{a}{\sqrt{c^2+d^2g_s^2}}dt\wedge dr -\frac{r^2
\eta}{H+r^2\eta^2} \frac{bg_s}{\sqrt{c^2+d^2g_s^2}}d\varphi\wedge
dz.
\end{eqnarray}
Using the Seiberg-Witten relation (\ref{2eq8}), we find the open
string moduli in this case: open string metric
\begin{equation}
\label{3eq13}
G_{ij} dx^i dx^j = h^{1/2}\sqrt{\frac{c^2+d^2g_s^2}{c^2+d^2g_s^2h}}(-dt^2
+dr^2 +r^2d\varphi^2 +dz^2),
\end{equation}
nonvanishing noncommutativity parameter
\begin{equation}
\label{3eq14}
\Theta^{tr}=-2\pi \alpha' \frac{c\eta r}{\sqrt{c^2+d^2g_s^2}}, \
\ \ \Theta^{\varphi z}=-2\pi \alpha'\frac{d\eta
g_s}{\sqrt{c^2+d^2g_s^2}},
\end{equation}
and open string coupling constant $G_s=\tilde g_s=
g_s^{-1}(c^2+d^2g_s^2)$.  In the decoupling limit (\ref{2eq12}),
we have
\begin{equation}
\label{3eq15} \alpha' G^{ij} \partial_i \partial_j = \frac{r \triangle
c}{\sqrt{c^2+d^2g_s^2}}\left(-\frac{d^2}{dt^2}+\frac{d^2}{dr^2}
+\frac{d^2}{r^2d \varphi^2}+\frac{d^2}{dz^2}\right),
\end{equation}
and
\begin{equation}
\label{3eq16}
\Theta^{tr}=-\frac{2\pi c \triangle r}{\sqrt{c^2+d^2g_s^2}}, \ \
\ \Theta^{\varphi z}=-\frac{2\pi \triangle d
g_s}{\sqrt{c^2+d^2g_s^2}}.
\end{equation}
We see from (\ref{3eq15}) and (\ref{3eq16}) that the decoupled
theory is an NCOS theory with both nonconstant time-space and
space-space noncommutativities and the effective open string scale
is $\alpha'_{\rm eff}= r\triangle c/\sqrt{c^2+d^2g_s^2}$ and open
string coupling constant $G_s=g_s^{-1}(c^2+d^2g_s^2)$. The
supergravity dual to this NCOS theory is
\begin{eqnarray}
\label{3eq17}
&& ds^2= \alpha' \tilde h^{-1/2} \frac{u^2}{\tilde
R^2}\sqrt{\frac{c^2+d^2g_s^2 \tilde h}{c^2+d^2g_s^2}} \times
\nonumber \\
&&~~~~~~~~~~ \left (
-\tilde f dt^2 +dr^2 +\tilde h( r^2d\varphi^2 +dz^2)
+ \frac{\tilde R ^4}{u^4}(\tilde f^{-1}du^2 +u^2
d\Omega_5^2)\right), \nonumber
\\
&& e^{-2\tilde \phi}= g_s^{2} \tilde h (c^2+d^2 g_s^2 \tilde
h)^{-2}, \ \ \ \tilde \chi =\frac{ac +db g_s^2 \tilde h}{c^2+d^2g_s^2
\tilde h}, \nonumber \\
&& 2\pi \alpha' B= \alpha' \frac{r^2\triangle u^4}{\tilde R^4
+r^2\triangle^2 u^4} \frac{dg_s}{\sqrt{c^2+d^2g_s^2}}d\varphi\wedge
dz -\alpha' \frac{\triangle r u^4}{\tilde R^4}
\frac{c}{\sqrt{c^2+d^2g_s^2}} dt\wedge dr,
\end{eqnarray}
where $\tilde R^4= 4\pi \tilde g_s N$. This supergravity
configuration (\ref{3eq17}) has the same Hawking temperature
(\ref{2eq19}) and entropy (\ref{2eq20}) as the configuration
(\ref{2eq17}) or (\ref{3eq9}). This implies that this set of
noncommutative theories (NCOS theory and NCSYM theory) has
completely the same thermodynamic properties in the dual supergravity
description.

\sect{Lorentz Boost and Light-like NCSYM}

In \cite{CO1} it is shown that starting from (space-like) NCSYM or (time-like)
NCOS theories, one can obtain light-like NCSYM theories by using Lorentz
transformation and rescaling some coordinates and variables. In
that case, all noncommutativity parameters are constant in those
theories. Here we show that the same approach gives us light-like
NCSYM theory with nonconstant noncommutativity parameter starting
from (space-like)NCSYM or (time-like) NCOS theory with nonconstant
noncommutativity parameter.

Let us begin with black D3-brane solution (\ref{2eq1}) in the Melvin universe
with $f=1$, namely, D3-brane solution with $H= 1+4\pi g_s
\alpha'^2/\rho^4$. Making a Lorentz boost along the coordinate $z$,
\begin{equation}
\label{4eq1}
z \to z \cosh\beta -t \sinh\beta, \ \ \ t\to t\cosh\beta
-z \sinh \beta,
\end{equation}
where $\beta$ is the boost parameter, and taking the limit
$\beta \to \infty$ and $\eta \to 0$ but keeping $\eta
\cosh\beta=\gamma$ as a constant, we obtain
\begin{eqnarray}
\label{4eq2}
&& ds^2= H^{-1/2}(-dz_+dz_- +dr^2 +r^2 d\varphi^2
-\frac{\gamma^2 r^2}{H}dz_-^2)+H^{1/2}(d\rho^2 +\rho^2d\Omega_5^2),
\nonumber \\
&& e^{2\phi}= g^2_s , \ \ \ 2\pi \alpha'B=- \gamma r^2 H^{-1}
d\varphi \wedge dz_-,  \nonumber \\
&& A_2=-g_s^{-1} H^{-1} \gamma  r  dz_-\wedge dr,\nonumber
\\
&& A_4=g_s^{-1} H^{-1} r  dt \wedge dr \wedge
d\varphi \wedge dz,
\end{eqnarray}
where $z_{\pm}= t \pm z$ are two null coordinates. This is a pp-wave solution
propagating along the direction $z$. But this is not a gravitational wave,
instead it is produced by the NSNS B-field. For this
background, using the Seiberg-Witten relation (\ref{2eq8}) we
obtain the open string moduli: open string metric
\begin{equation}
\label{4eq3}
G_{ij} dx^i dx^j = -dz_+dz_- +dr^2 +r^2d\varphi^2,
\end{equation}
nonvanishing noncommutativity parameter
\begin{equation}
\label{4eq4}
\Theta^{z_+\varphi}= 4\pi \alpha' \gamma,
\end{equation}
and open string coupling constant $G_s=g_s$. In the decoupling limit,
\begin{equation}
\label{4eq5}
\alpha' \to 0, \ \ \gamma = \Gamma /\alpha',
\end{equation}
and keeping $g_s$ to be constant, the decoupled theory
on the D3-branes is a light-like NCSYM theory living in
(\ref{4eq3}) with the Yang-Mills coupling $g^2_{\rm YM} =
2\pi g_s$ and nonconstant noncommutativity parameter
$\Theta^{z_+\phi}= 4\pi \Gamma$.  That the noncommutativity parameter
is not a constant can be seen easily in the coordinates $(x,y,z)$:
\begin{equation}
\Theta_{z_-x}= -2\pi \Gamma y, \ \ \ \Theta_{z_-y}=2\pi \Gamma x.
\end{equation}
The dual supergravity  description can be obtained from the
solution (\ref{4eq2}) by taking the decoupling limit (\ref{4eq5})
and $\rho = \alpha' u$. The geometry configuration is
\begin{eqnarray}
\label{4eq7}
&& ds^2=\alpha' \frac{u^2}{R^2}\left( -dz_+dz_-+dr^2
+r^2d\varphi^2-\frac{\Gamma^2 r^2 u^4}{R^4} dz_-^2
+\frac{R^4}{u^4}(du^2 +u^2d\Omega_5^2)\right), \nonumber
\\
&& e^{2\phi}=g_s^2, \ \ \ 2\pi \alpha' B= -\alpha' \Gamma r^2
\frac{u^4}{R^4} d\varphi \wedge dz_-, \nonumber \\
&& A_2= -\alpha'g_s^{-1}\Gamma r \frac{u^4}{R^4} dz_-\wedge dr,
\nonumber \\
&& A_4 = \alpha'^2 g_s^{-1}r \frac{u^4}{R^4} dt\wedge dr
 \wedge d\varphi\wedge dz,
\end{eqnarray}
where $R^4= 4\pi g_sN$. From the supergravity configuration we can
see that near the origin, $u^4\ll R^4/\Gamma^2 r^2$,  the
deviation from the $AdS_5\times S^5$ is small. This implies that
in the low-energy limit, the difference of the light-like NCSYM
theory with nonconstant noncommutativity parameter from ${\cal N}=4$ SYM
theory is negligible.

It is easy to see that we can obtain the same light-like NCSYM
theory if we start from the supergravity dual (\ref{3eq1}) or more
general (\ref{3eq12}) for NCOS theory with nonconstant
noncommutativity parameter. We will not repeat to produce the
result. Instead we will consider the black D3-brane solution
(\ref{2eq1}), namely, nonextremal case, in the Melvin universe. In
this case, $f=1-\rho_0^4/\rho^4$. Applying the Lorentz boost
(\ref{4eq1}) to the solution (\ref{2eq1}), and taking the limit,
$\eta \to 0$, $\alpha \to \infty$, $\beta \to \infty$, but keeping
$\eta \cosh\beta =\gamma $, $\rho_0^4 \cosh\alpha \sinh\alpha =
4\pi g_s\alpha'^2 N$ and $\rho_0^4\cosh^2\beta =P$ as constant, we
have solution
\begin{eqnarray}
\label{4eq8} && ds^2= H^{-1/2}(-dz_+dz_- +dr^2 +r^2 d\varphi^2
-\frac{\gamma^2 r^2}{H}dz_-^2 +\frac{P}{\rho^4}dz_-^2)
+H^{1/2}(d\rho^2 +\rho^2d\Omega_5^2),
\nonumber \\
&& e^{2\phi}= g^2_s , \ \ \
2\pi \alpha'B=- \gamma r^2 H^{-1} d\varphi \wedge dz_-,  \nonumber \\
&& A_2=-g_s^{-1} H^{-1} \gamma  r  dz_-\wedge dr,\nonumber
\\
&& A_4=g_s^{-1} H^{-1} r dt \wedge dr \wedge
d\varphi \wedge dz,
\end{eqnarray}
where $P$ has an interpretation as momentum of gravitational
pp-wave propagating along direction $z$. It is easy to show that
the decoupling limit of open string ending on D3-branes in the
background (\ref{4eq8}) is the same as that of the background
(\ref{4eq2}), that is, it is a light-like NCSYM theory with
nonconstant noncommutativity parameter. Due to the appearance
of the momentum $P$, the field theory  lives in the background
(\ref{4eq3}), but in an infinite-momentum frame~\cite{CO1,ppv}.
The supergravity dual is obtained by taking decoupling limit
(\ref{4eq5}) combining with
\begin{equation}
\label{4eq9}
\rho= \alpha' u, \ \ \ P= \alpha'^4 \tilde P,
\end{equation}
with $\tilde P$ being a constant. The supergravity configuration is
\begin{eqnarray}
\label{4eq10}
&& ds^2=\alpha' \frac{u^2}{R^2}\left( -dz_+dz_-+dr^2
+r^2d\varphi^2-\frac{\Gamma^2 r^2 u^4}{R^4} dz_-^2 +\frac{\tilde
P}{u^4}dz_-^2
+\frac{R^4}{u^4}(du^2 +u^2d\Omega_5^2)\right), \nonumber
\\
&& e^{2\phi}=g_s^2, \ \ \ 2\pi \alpha' B= -\alpha' \Gamma r^2
\frac{u^4}{R^4} d\varphi \wedge dz_-, \nonumber \\
&& A_2= -\alpha'g_s^{-1}\Gamma r \frac{u^4}{R^4} dz_-\wedge dr,
\nonumber \\
&& A_4 = \alpha'^2 g_s^{-1}r \frac{u^4}{R^4} dt\wedge dr
\wedge d\varphi\wedge dz,
\end{eqnarray}
where $R^4=4\pi g_s N=2 g^2_{\rm YM}N$. When $\tilde P=0$, the
configuration (\ref{4eq10}) reduces to (\ref{4eq7}), which is the
supergravity configuration dual to the light-like NCSYM theory with
nonconstant noncommutativity in a static frame.

\sect{Conclusion}

According to the holographic principle, quantum theory of gravity
must be a nonlocal quantum theory. Therefore studying nonlocal
field theory is of great interest in its own right. More recently
Hashimoto and Thomas~\cite{HT} have shown that the decoupling
theory on D3-branes in Melvin universe supported by a space-like
NSNS B-field is an NCSYM theory with  space-dependent space-space
noncommutativity. This generalized the case of the NCSYM theory
with  constant space-space noncommutativity~\cite{NCYM}. In this
paper we have first extended the supergravity dual to the NCSYM
theory with space-dependent noncommutativity to the nonextremal
case, which describes the NCSYM theory at finite temperature. We
have shown that under S-duality, the NCSYM theory with
space-dependent noncommutativity changes to be an NCOS theory with
space-dependent space-time noncommutativity, while it is an NCOS
theory with both space-dependent space-space and space-time
noncommutativities for a general $SL(2,\mathbb{Z})$
transformation. These NCOS theories with space-dependent
noncommutativity widen the family of NCOS theories with constant
noncommutativity parameter~\cite{NCOS}. Furthermore we have found
that in the dual supergravity description, these NCSYM and NCOS
theories have completely the same thermodynamics as that of
ordinary SYM theory, NCSYM and NCOS theories with constant
noncommutativity. This implies that these theories have the same
degrees of freedom in the dual supergravity description.  In
addition, starting from black D3-brane solution in Melvin universe
and making Lorentz boost along a spatial direction on the
worldvolume, we have shown that the decoupled theory is a
light-like NCSYM theory with space-dependent noncommutativity in
static frame or infinite-momentum frame, depending on whether
there is a gravitational pp-wave on the worldvolume of D3-branes.

\section*{Acknowledgments}
This work was supported in part by a grant from Chinese Academy of
Sciences, and grants from NSFC, China (No. 10325525 and No.
90403029), and by the Grant-in-Aid for Scientific Research Fund of
the JSPS No. 16540250.

\end{document}